\documentclass[11pt]{article}

\columnwidth\textwidth

\begin{document}

\title{Fock Vacuum Instability and Causality}

\author{Andreas Aste\\ 
Institut f\"ur Theoretische Physik der Universit\"at Basel,
Klingelbergstrasse 82, 4056 Basel\\
Switzerland}
\date{April 2, 2001}

\maketitle
 
%\newpage

\begin{abstract}
The vacuum diagram is calculated at second order for
theories with self-interacting massless fields
in the framework of finite causal perturbation theory. It is
pointed out that the infrared behaviour of the vacuum diagram leads
to unstable Fock vacua for QCD or massless QED, but not for
quantum gravity.
Therefore a radical rearrangement of the physical system
must take place for such theories. Conversely, stability of the
Fock vacuum for massless interacting fields
is another hint at the possibility that
quantum gravity should be treated as an effective theory.
\end{abstract}

\newpage
%\twocolumn
%\setlength{\baselineskip}{22pt}

\section{Introduction}
Soon after the beginnings of Quantum Field Theory (QFT), it turned out that the
perturbative expansion of several quantities like the electron self-energy
are plagued by divergences of different kind. These divergences
in QFT can be classified into four types, namely infrared, ultraviolet (UV),
infinite volume and particle number divergences.
A large amount of work has
been done to overcome these problems, but even after the construction
of a renomalized perturbation expansions based on the Lagrangean formalism
for interacting fields or the scattering matrix
($S$-matrix), little is known until today about the
convergence properties of the perturbation expansion.

The first attempt to go beyond the widely used
Lagrangean framework can be traced
back to W. Heisenberg in 1943 \cite{heisenberg},
where he came to the conclusion that the
basic observable is the $S$-matrix, which describes actually the
quantities that can be measured in an experiment. Heisenberg even proposed
to construct a theory directly in terms of the elements of the $S$-matrix,
eliminating the notion of fields and the hypothesis of adiabatic switching
of the interaction. However, the complete banishing of local quantities from
the theory turned out to be too radical.

A very natural and rigorous approach to axiomatic perturbation theory
which has been applied successfully to all relevant interactions of the
Standard model
\cite{appl1,appl2,appl3}
was provided by H. Epstein and V. Glaser in 1971 \cite{eg,eg2,scharf1}.
Their method, often called
finite causal perturbation theory or FCPT for short in this paper,
avoids UV-divergences from the start by
defining mathematically correct time-ordered products for distributions.
The resulting distributions are
smeared out by test functions, avoiding provisionally volume and infrared
divergences at once. 
In FCPT,
the ansatz for the $S$-matrix as a power 
series in the coupling constant is crucial, namely $S$ is considered as a sum of 
smeared operator-valued distributions
\begin{equation}
S(g)={\bf 1}+\sum_{n=1}^{\infty}\frac{1}{n!} \int\!\! d^{4}x_{1}
\ldots d^{4}x_{n}\, T_{n}(x_{1},\ldots, x_{n})\,g(x_{1})\cdot\ldots\cdot 
g(x_{n})\quad  ,
\end{equation}
where the  Schwartz test function $g\in\cal{S}(\bf{R}^{4})$ switches the 
interaction and provides the infrared cutoff.
The basic formulation of causality in FCPT, which had already been
used by Bogoliubov {\em{et al.}} \cite{bogol}, is
\begin{equation}
S(g_1+g_2)=S(g_2)S(g_1) \quad , \label{causality1}
\end{equation}
if the support of $g_2(x)$ is later than $\mbox{supp} \, g_1$ in some
Lorentz frame. We denote this by $\mbox{supp} \, g_2 > \mbox{supp} \, g_1$.
The condition allows the construction of
the $n$-point distributions $T_n$ as a  well-defined  `renormalized'
time-ordered product expressed 
in terms of Wick monomials of free fields $:\!{\cal O}(x_{1},\ldots,x_{n})$:
\begin{equation}
T_{n}(x_{1},\ldots, x_{n})=\sum_{ { \cal{ O}}} :\!{\cal 
O}(x_{1},\ldots,x_{n})\!:\,
t_{n}^{{ \cal O}}(x_{1}-x_{n},\ldots,x_{n-1}-x_{n})\,. \label{smatrix}
\end{equation}
Possible ambiguities in the decomposition of $T_n$ according to
({\ref{smatrix}}) will play no role for us in the sequel.
The $t_{n}$ are C-number distributions. $T_{n}$ is constructed inductively from 
the first order $T_{1}(x)$,
which describes the interaction among the quantum fields,  and from 
the lower orders $T_{j}$, $j=2,\ldots,n-1$ by means of Poincar\'e
covariance and causality.
The inductive construction of the n-point distributions $T_n$
can be considered
as the main technical drawback of the theory, since all lower orders
$T_1,...T_{n-1}$ must be calculated first in order to construct $T_n$.
But a naive definition of $T_n$ like
\begin{displaymath}
T_n(x_1,...,x_n)=
\end{displaymath}
\begin{equation}
\sum_{\pi} \Theta(x_{\pi_1}^0-x_{\pi_2}^0)\cdot ...
\cdot \Theta(x_{\pi_{n-1}}^0-x_{\pi_n}^0) T_1(x_{\pi_1})\cdot ... \cdot
T_1(x_{\pi_n})
\end{equation}
where the sum runs over all $n!$ permutations,
is not well-defined, since it contains the product of Heaviside
distributions with other singular distributions. This error leads to the
well-known UV divergences in the calculation of Feynman diagrams.
We illustrate this fact by the following very simple example in one
dimension:
It is of course meaningless to multiply a Heaviside
distribution $\Theta(t)$ with
a $\delta$-distribution $\delta(t)$ or even its
derivative $\delta'(t)$. The Fourier transforms are
\begin{equation}
\hat{\Theta}(\omega)=-\frac{i}{\sqrt{2 \pi}} \frac{1}{\omega-i0} \quad ,
\quad \hat{\delta'}(\omega)=\frac{i\omega}{\sqrt{2 \pi}} \quad .
\end{equation}
The ill-defined product $(\Theta \delta')(t)$ goes over into a 
non-existing convolution by Fourier transform
\begin{equation}
(\Theta\delta')(t) \rightarrow \frac{1}{2 \pi} (\hat\Theta \ast
\delta')(\omega)=\frac{1}{(2 \pi)^2} \int_{-\infty}^{+\infty}
d\omega' \, \frac{\omega-\omega'}
{\omega'-i0}
\end{equation}
which is definitely 'UV-divergent'.

$T_2(x,y)$ contains typically
tree (one contraction), loop (two contractions) and vacuum graph (three 
contractions)
contributions. Due to the presence of normal ordering, tadpole diagrams do not 
show up. In this paper, we will concentrate on the
vacuum graph, which has been considered already in \cite{nuovoc} in a
preliminary way.
Since this graph contains no external fields, we can directly
identify $T^{vac}_2(x_1,x_2)=t^{vac}_2(x_1-x_2)$. Usually, vacuum bubbles
are neglected by simply absorbing them into a phase of the $S$-matrix. In
FCPT, this is not necessary, since their contribution to the $S$-matrix
vanishes in the adiabatic limit for theories like ordinary QED with a
massive electron. But this is not always the case for theories with
self-interacting massless fields like QCD.

\section{Preliminary remarks}
The first order interaction $T_1(x) \sim i:{\cal{L}}_{int}:$ is
usually motivated from a classical Lagrangean interaction density.
For QED, $T_1(x)$ is given by means of the free electron and photon field
\begin{equation}
T_1(x)=ie:\overline{\Psi}(x) \gamma^\mu \Psi(x): A_\mu(x) \quad .
\end{equation}
For the purpose of this paper, we outline briefly the main steps in the 
construction of $T_{2}(x,y)$ from 
a given  first order interaction.
In FCPT, one constructs first the operator valued distributions
\begin{eqnarray}
R'_2(x,y) &=&-T_1(y)T_1(x) \quad , \\
A'_2(x,y) &=&-T_1(x)T_1(y) \quad , \quad \mbox{and} \\
D_2(x,y)&=&R'_2(x,y)-A'_2(x,y)=[T_1(x),T_1(y)] \quad .
\end{eqnarray}
A meaningful theory ensures causality conditions (in a distributive
sense) like (\ref{causality1}), which implies also
\begin{equation}
[T_1(x),T_1(y)]=0 \quad \mbox{for} \quad (x-y)^2<0
\end{equation}
such that
\begin{equation}
\mbox{supp} \, D_2(x,y) \subseteq \{(x,y) \in {\bf{R}}^8 | (x-y)^2 \geq 0 \}
=\overline{V}_2
\quad .
\end{equation}
This property of $D_2$ must be checked first before one may start the inductive
construction of the $S$-matrix.
From $D_2(x,y)$ we construct the Lorentz covariant retarded distribution
$R_2(x,y)$, which coincides with $D_2(x,y)$ in the sense that
$(R_2, \varphi)=(D_2, \varphi)$ for all test functions
$\varphi \in {\cal{S}}({\bf R}^8)$ with (compact)
$\mbox{supp} \, \varphi \subset \{(x,y) \in {\bf R}^8 | (x^0-y^0) > 0\}$
and $(R_2,\varphi)=0$ for all $\varphi$ with $\mbox{supp} \, \varphi \subset
\{ (x,y) \in {\bf R}^8 | (x^0-y^0) < 0 \}$.
Such a distribution can always be found, but it is only defined up to
local operator-valued distributions with support on the complete diagonal
$\Delta_2=\{(x,y) \in {\bf{R}}^8 | x=y\}$. Symmetry and renormalizability
considerations further restrict these ambiguous terms.
Obviously, $T_2(x,y)=R_2(x,y)-R'_2(x,y)$ is then a time-ordered product,
since by construction
\begin{equation}
T_2(x,y)=\left\{ \begin{array}{r@{\quad:\quad}l}
T_1(x)T_1(y) & x^0-y^0>0 \\
T_1(y)T_1(x) & x^0-y^0 < 0
\end{array} \right. \quad .
\end{equation}
We pursue the strategy described above for the calculation of the vacuum
diagram described by the C-number distribution $t_2^{vac}(x-y)$. 

\section{Calculation of the vacuum diagram}
We calculate the vacuum diagram for the example of pure QCD,
where the first order self-interaction is described by the
help of $SU(3)$ structure constants $f_{abc}$
\begin{eqnarray}
T_1(x)&=&igf_{abc}:A_{\mu a}(x) A_{\nu b}(x) \partial^\nu A^\mu_c(x): \nonumber
\\
& & -igf_{abc}:A_{\mu a}(x) u_b(x) \partial^\mu \tilde{u}_c(x): \quad .
\end{eqnarray}
Here, $A_{\mu a}$ are the free gauge potentials, satisfying the
commutation relations
\begin{equation}
[A_a^{\mu(\mp)}(x),A_b^{\nu(\pm)}(y)]=i\delta_{ab}g^{\mu \nu} D_0^{\pm}(x-y)
\quad ,
\end{equation}
where $A^{(\pm)}$ are the emission and absorption parts of $A$ and
$D_0^{\pm}$ the zero-mass Pauli-Jordan distributions defined by
\begin{equation}
D_0^{\pm}(x)=\pm \frac{i}{(2 \pi)^3} \int d^4 p \, \Theta(\pm p^0)
\delta(p^2) e^{-ipx} \quad .
\end{equation}
The free ghosts are Fermi fields and satisfy the anticommutation relations
\begin{equation}
\{u_a^{(\mp)}(x),\tilde{u}_b^{(\pm)}(y) \} = -i \delta_{ab}
D_0^{\pm}(x-y) \quad ,
\end{equation}
and all other anticommutators vanish. 
The quartic gluon interaction is missing in $T_1$ since it is enforced
by gauge invariance as a local counterterm at second order
in FCPT \cite{appl1}.
For the sake of convenience, we will use $g$ for the QCD coupling constant
as well as for test functions, since a mix up is impossible.

As described in the previous section,  we
calculate first the vacuum expectation value
\begin{equation}
R_2^{'vac}(x,y)=r'^{vac}_2(x-y)=-<\Omega | T_1(y)T_1(x) | \Omega> \quad.
\end{equation}
The calculation is straightforward, therefore we present the result
which can be decomposed in a purely gluonic part and a ghost part:
\begin{equation}
r'^{vac,gl}_2(x-y)=-3ig^2N(N^2-1)D_0^-(x-y) \partial_\nu^x D_0^-(x-y)
\partial_y^\nu D_0^-(x-y) \quad ,
\end{equation}
\begin{equation}
r'^{vac,gh}_2(x-y)=+ig^2N(N^2-1)D_0^-(x-y) \partial_\nu^x D_0^-(x-y)
\partial_y^\nu D_0^-(x-y) \quad ,
\end{equation}
where we have used that the fully antisymmetric structure constants satisfy
$f_{abc}f_{abc}=N(N^2-1)$ for $SU(N)$.

It is difficult to give a meaning to the product of three singular objects
in x-space, but $r'^{vac}_2$ is of course well-defined via the impulse space,
where the products in x-space go over into convolutions.
The C-number distribution we consider now is therefore
(omitting temporarily some prefactors)
\begin{equation}
r'^{vac}_2(x)=-D_0^-(x) \partial_\mu D_0^-(x) \partial^\mu D_0^-(x)=
-\frac{1}{6} \partial_\mu \partial^\mu (D_0^{-}(x))^3 \quad ,
\end{equation}
\begin{equation}
\hat{r}'^{vac}_2(p)=\frac{p^2}{6(2 \pi)^4} (\hat{D}^-_0 \ast \hat{D}^-_0
\ast \hat{D}^-_0)(p) \quad .
\end{equation}
The preliminary result
\begin{equation}
 (\hat{D}^-_0 \ast \hat{D}^-_0)(p)=-\frac{1}{8 \pi} \Theta(p^2) \Theta(-p^0)
\end{equation}
can be readily obtained. Then we have
\begin{equation}
\hat{r}'^{vac}_2(p)=\frac{i p^2}{24(2 \pi)^6} \int d^4q \, \delta((p-q)^2)
\Theta(q^0-p^0) \Theta(q^2) \Theta(-q^0) \quad . \label{integral}
\end{equation}
If $p$ is spacelike, i.e.\ $p^2<0$, then we can choose a special Lorentz frame
where $p=(0,\vec{p})$, and it follows that
$\hat{r}'^{vac}_2(p)=0$ due to the term
$\Theta(q^0) \Theta(-q^0)$ in (\ref{integral}).
For timelike $p$ we choose a Lorentz frame where $p=(p^0,\vec{0})$.
The four-dimensional integral can be restricted to three dimensions
because of the $\delta$-distribution in $(p-q)^2$.
Taking the Heaviside distributions in (\ref{integral}) into account,
\begin{equation}
(p-q)^2=p_0^2-2p^0q^0+q_0^2-\vec{q}^{\,2}=0
\end{equation}
implies $q^0=p^0+|\vec{q}|$ and the integral
goes over into
\begin{eqnarray}
\hat{r}'^{vac}_2(p^0,\vec{0})&=&\frac{ip_0^2}{24(2 \pi)^6} \int
\frac{d^3q}{2 |\vec{q}|} \Theta(p_0^2+2p^0 |\vec{q}|)
\Theta(-|\vec{q}|-p^0) \nonumber \\
&=&\frac{ip_0^2}{24 (2 \pi)^5} \Theta(-p^0) \int_0^{-p^0/2}
d|\vec{q}| |\vec{q}| \nonumber \\
&=&\frac{ip_0^2}{192 (2 \pi)^5} \Theta(-p^0) p_0^2 \quad .
\end{eqnarray}
The final result is by Lorentz covariance
\begin{equation}
\hat{r}'^{vac}_2(p)=\frac{i}{192 (2 \pi)^5} \Theta(-p^0) \Theta(p^2) (p^2)^2 
\quad .
\end{equation}
From symmetry considerations we have automatically
\begin{equation}
\hat{a}'^{vac}_2(p)=\hat{r}'^{vac}_2(-p)
\end{equation}
and therefore
\begin{equation}
\hat{d}^{vac}_2(p)=-\frac{i}{192 (2 \pi)^5} \mbox{sgn}(p^0) \Theta(p^2) (p^2)^2 
\quad .
\end{equation}
The calculation of the retarded part $r^{vac}_2$ of $d^{vac}_2$ is demonstrated
for an analogous case in \cite{ym2} in full detail.
The result turns out to be
\begin{equation}
\hat{r}^{vac}_2(p)=\frac{1}{192 (2 \pi)^6} (p^2)^2
\log \frac{-(p^2+ip^00)}{M^2}
\quad ,
\end{equation}
and the final result for the vacuum diagram is
\begin{equation}
\hat{t}^{vac}_2(p)=-\frac{ig^2N(N^2-1)}{96 (2 \pi)^6} (p^2)^2 \log
\frac{-(p^2+i0)}{M^2}
\quad .
\end{equation}
Scale invariance of the causal distribution is destroyed by the
time-ordering procedure.
The parameter $M^2$ remains undefined. Changing the value of $M^2$ changes
$t^{vac}_2$ only by local terms in x-space,
which are not relevant for the following discussion.

\section{Volume divergence}
A problem arises when we calculate the vacuum transition amplitude
at second order
\begin{eqnarray}
< \Omega | S_2(\Omega) | \Omega> &=& \frac{1}{2}
\int d^4x d^4y \, g(x)t_2^{vac}(x-y)g(y) \nonumber \\
&=& \frac{1}{2}(2 \pi)^2 \int d^4p \, \hat{g}(-p) \hat{t}_2^{vac}(p) \hat{g}
(p)
\end{eqnarray}
in the adiabatic limit $g \rightarrow 1$. Since $1 \not\in
{\cal{S}}({\bf{R}}^4)$, we have to specify the limiting process.
Usually, one chooses a fixed test function $g_0 \in {\cal{S}}({\bf{R}}^4)$
with $g_0(0)=1$ and performs a scaling limit, i.e. \ for $\varepsilon
\rightarrow 0$ we have $g_\varepsilon(x)=g_0(\varepsilon x) \rightarrow 1$.
This implies
\begin{eqnarray}
< \Omega | S_2(\Omega) | \Omega> &=& \frac{1}{2}
\int d^4x d^4y \, g_0(\varepsilon x)t_2^{vac}(x-y)g_0(\varepsilon y)
\nonumber \\
&=& \frac{1}{2 \varepsilon^8} \int d^4xd^4y \,
g_0(x) t_2^{vac}((x-y)/\varepsilon) g_0(y) \nonumber \\
&=&\frac{(2 \pi)^2}{2 \varepsilon^4} \int d^4p \, \hat{g}_0(-p)
\hat{t}_2^{vac}(\varepsilon p) \hat{g}_0(p) \quad .
\end{eqnarray}
This expression diverges in the adiabatic limit, since
\begin{equation}
\frac{1}{\varepsilon^4}\int d^4p \, \hat{g}_0(-p) (\varepsilon p)^4
\log \frac{-(\varepsilon^2p^2+i0)}{M^2} \hat{g}_0(p) \rightarrow
\end{equation}
\begin{equation}
\int d^4p \, \hat{g}_0(-p) p^4
\log \frac{-(p^2+i0)}{M^2} \hat{g}_0(p)
+\log(\varepsilon^2) \int d^4p \, \hat{g}_0(p) p^4
\hat{g}_0(-p) \label{divexp}
\end{equation}
due to the logarithmic term in $\varepsilon$.

\section{Conclusion}
The Fock vacuum is unstable against the interaction
of the massless gluon fields. This is a consequence of the factor
$p^4$ in the vacuum diagram, which expresses simply its
power-counting degree $\omega =4$. 
The instability has nothing to do with the non-abelian character
of QCD, since the same phenomenon also occurs in massless
QED. There, $t_2^{vac}$ has exactly the same form as in QCD
up to a negative prefactor
\begin{equation}
\hat{t}_2^{vac}=\frac{ie^2}{24 (2 \pi)^6} p^4 \log \frac{-(p^2+i0)}{M^{'2}}
\quad .
\end{equation}
It is tempting to look for a cancellation of the QCD vacuum divergence
through massless fermions, but in view of the present properties of QCD
(confinement), this is probably not even desirable.

It is interesting to observe
that in quantum gravity, the situation is remedied by
the non-renormalizable coupling that leads to a factor $p^6$
in $\hat{t}_2$ which cancels
the logarithmic divergence. The same is true for the Euler effective
action for photon-photon scattering
\begin{equation}
T_1(x)=\frac{2i\alpha^2}{45m_e^4}[(\vec{E}^2-\vec{B}^2)^2+7(\vec{E}
\vec{B})^2] \quad .
\end{equation}
One might therefore argue for QCD that the volume divergence encountered in
eq. (\ref{divexp})
must be compensated by the interaction, which signals a radical rearrangement
of the system as in statistical mechanics. Such a rearrangement is
well-known from (1+1)-dimensional QFT models, especially for the
Schwinger model \cite{schw1,schw2,schw3}.

Our result shows also that it is really a fair
assumption that quantum gravity should be treated as
an effective theory \cite{donoghue}, which incorporates the 'true'
physical vacuum. The price one has to pay for perturbative vacuum stability
is non-renormalizability,
since we do not know the underlying theory of quantum gravity.

\end{document}